\documentstyle[aps,prl,amssymb,rotating,epsfig,color,graphics,eepic,twocolumn]{revtex}
\begin{document}
\twocolumn[\hsize\textwidth\columnwidth\hsize\csname @twocolumnfalse\endcsname
\title{Force Dependence of the Michaelis Constant in a Two--State
Ratchet Model for Molecular Motors}
\author{G Lattanzi$^*$ \and A Maritan$^{*,\dagger}$}
\address{E--mail: lattanzi@sissa.it}
\address{$^*$ International School for Advanced Studies and 
Istituto Nazionale di Fisica della Materia, \\
Via Beirut 2-4, 34014 Trieste, Italy and \\ 
$^\dagger$ The Abdus Salam International Center for Theoretical Physics, \\
Strada Costiera 11, 34100 Trieste, Italy.}
\date{\today}

\bibliographystyle{unsrt}

\maketitle

\begin{abstract}

We present a quantitative analysis of recent data on the kinetics of ATP hydrolysis, 
which has presented a puzzle regarding the load dependence of the Michaelis constant. 
Within the framework of coarse grained two--state ratchet models, our analysis not 
only explains the puzzling data, but provides a modified Michaelis law, which could be 
useful as a guide for future experiments.


\end{abstract}
]

Molecular motors or protein motors are the terms used to describe a highly 
specialized class of enzymes which transduce the energy excess in the chemical 
hydrolysis reaction of ATP (adenosinetriphosphate) into mechanical work. 
They are involved in several important cellular processes, ranging from the 
transport of material and vesicles, to cell mobility and cell division.
Linear protein motors move along complex periodic and polar structures called
filaments, obtained by the polymerization of a single monomer (actin) or dimer
(microtubules)~\cite{bio,hux67}.

Several models have been proposed so far to explain the energy
transduction process.
In the earliest ones, known as cross--bridge models~\cite{hux57,hill74},
the motor can exist in different states (up to 5 or
6~\cite{tho98}), within each of which the system reaches local thermodynamic 
equilibrium on time scales small compared to the exchange rates between 
the states. This hypothesis relies on the fact that the characteristic
times of the motors (measured through transient response~\cite{hux71}) are of the 
order of  milliseconds, while thermal equilibrium on a length scale of about 
10 nm occurs in 10-100 ns~\cite{leib93,par99}. The latest 
models~\cite{mag93,pcpa94,doer94,jp95,der95,ast97,kol98} share some common 
features:
i) need for asymmetry (polarity) in order to establish a certain preferred
direction of motion;
ii) chemical energy consumption as a source of mechanical work; iii) 
local thermodynamic equilibrium for each state.
A particular promising approach is that due to 
Vale and Oosawa~\cite{vale90}, using the ratchet
concept introduced  by Feynman~\cite{fey66}. Since then, ratchet models 
have been intensively studied~\cite{mag93,pcpa94,jap97}.
Non--equilibrium thermodynamics as well as stochastic methods contribute
to the theoretical background of this new field of research. 

The role of force in the reaction kinetics is still an open
question,~\cite{duk96,fish99,schn99}. In particular, recent experiments performed 
on kinesin (a processive motor),~\cite{schn99}, seem to indicate that the 
kinetics of ATP hydrolysis can be described by the Michaelis--Menten mechanism. This
simple mechanism, introduced in 1913 to describe the process of enzymatic 
catalysis~\cite{fersht}, assumes that the catalytic reaction of a substrate $S$
is divided into two processes. The enzyme $E$ and the substrate first combine
rapidly and reversibly to give an enzyme--substrate complex $ES$ with no
chemical change on the substrate. The chemical reaction occurs in a
second step with a first--order rate constant $k_{cat}$, or turnover number.
It is then simple to show that the reaction velocity $r$ (or rate of $S$
consumption) is given by the Michaelis law:

\begin{equation}
r=\frac{r_{max}\ [S]}{K_M + [S]},
\label{eq:mm}
\end{equation}

\noindent where $r_{max}$ is a saturation value and $K_M$ is called the Michaelis
constant. For the ATP hydrolysis reaction, $[S]$ is simply replaced by 
$[ATP]$. Furthermore it is reasonable that the velocity of a molecular motor 
should depend  strictly on the rate of ATP consumption~\cite{leib93}, so that the
velocity curve is given by a Michaelis law in terms of ATP concentration. This
hypothesis has been tested in the experiment~\cite{schn99} at least for high ATP 
concentrations. According to these experimental findings the Michaelis law 
can be written as:

\begin{equation}
v = \frac{\epsilon \ p \ k_{cat} \ [ATP]}{K_M+[ATP]},
\label{eq:exp}
\end{equation}

\noindent where $p$ is the periodicity of the filament, $k_{cat}$ is the
turnover number, and $\epsilon$, the coupling ratio, gives an 
estimate of the average performance of one hydrolysis event, i.e. it is higher 
when the energy transduction process is more efficient.
It is rather intuitive that the coupling ratio should depend on the applied 
force, and should decrease when the opposing force is increased, as 
the applied load strongly limits the maximum attained velocity on which the 
coupling ratio is linearly dependent.
The rather surprising experimental finding~\cite{schn99} is that 
the Michaelis constant $K_M$ is no longer a constant, but increases with increasing 
applied load. 
This result is not intuitive and rather striking, since the 
Michaelis constant is just an equilibrium constant for the reaction leading to 
the formation of the enzyme-substrate complex. Indeed reaction rates are 
assumed to be independent of the external force as long as they do not 
contribute to a net motion of the motor. Thus this experimental finding 
makes the 
picture more complicated than expected. It was suggested in~\cite{schn99} that a
possible explanation might be
to insert the external force in the transition rates as 
done by Fisher and Kolomeisky~\cite{fish99}, but this cannot be done
unambiguously, as stated by the authors themselves.

\begin{figure}
\begin{center}
\epsfig{figure=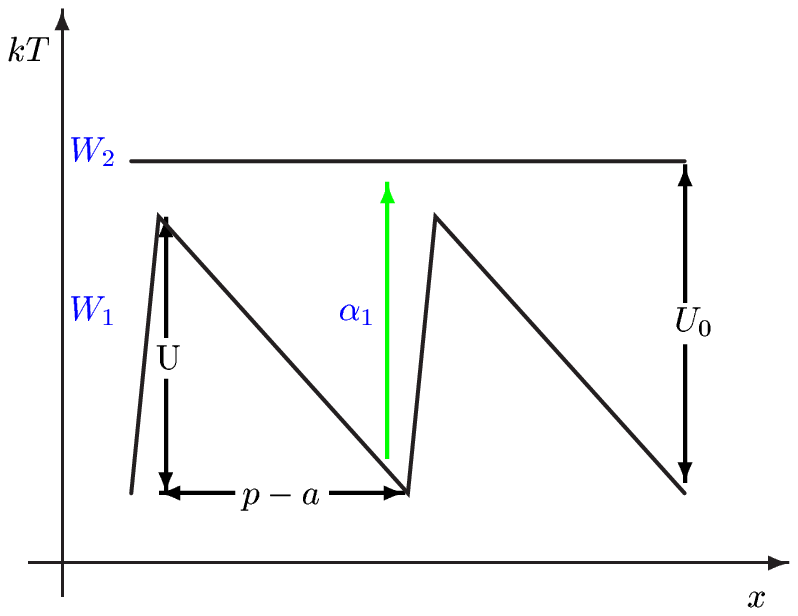,width=7cm}
\caption[Figure~\ref{fig:model}]{Model $(a)$: ratchet potential ($W_1$) and 
diffusive state $2$. The parameters are chosen so that $a/p=0.1$, $U=10 kT, 
U_0=12 kT, \omega/ D = 50$.}
\label{fig:model}
\end{center}
\end{figure}

Coarse--grained two-state ratchet models~\cite{jap97} can give further insight 
into this problem. In these models, the state of the motor is indicated 
by the index $i$, whereas $x$ is the  position of the protein center of mass 
along the track. 
The chemical reactions force the motor protein to switch from one 
state $i$ at position $x$ to another state $j$ in the same position $x$, 
with a rate given by $\omega_{ij}(x)$. 
The motor moves under the influence of a potential 
$W_i(x)$ chosen so as to reproduce the interaction between the filament 
and the motor head and with the same characteristics as the filament: 
polarity and periodicity. For this reason the potential is chosen to 
be asymmetric and periodic with period $p$ for each state $i$  (``ratchet''
potential, see Fig.~\ref{fig:model}). 

In a Fokker--Planck~(FP) description, $P_i(x,t)$  is the  probability density 
for the particle to be in state $i$ at position $x$ at time $t$, while 
the probability current $J_i(x,t)$ is defined as:

\begin{equation}
J_i(x,t) = -\frac{D_i}{kT} \left[ P_i(x)\partial_x W_i -
 P_i(x) F_{ext} + kT \partial_x P_i(x) \right]
\label{eq:fp}
\end{equation}

\noindent
where $D_i$ is a diffusion coefficient which is taken to be equal in both 
states, $T$ is the temperature, $k$ is the Boltzmann constant and $F_{ext}$ is 
the  external force. Since we are dealing with a simple two-state model, the 
chemical reaction cycle is compressed in a two step process:

\renewcommand{\arraystretch}{0.1}
\begin{equation}
\begin{array}{ccc}
 & {\scriptstyle \alpha_1(x)} &  \\
M + ATP  & {\textstyle \rightleftharpoons} & M\cdot ADP \cdot P \\
 & {\scriptstyle \alpha_2(x)} & 
\end{array}
\end{equation}
\begin{equation}
\begin{array}{ccc}
 & {\scriptstyle \beta_1(x)} &  \\
M + ADP + P & {\textstyle \rightleftharpoons} & M\cdot ADP \cdot P \\
 & {\scriptstyle \beta_2(x)} &
\end{array}
\label{eq:adp}
\end{equation}

\renewcommand{\arraystretch}{1.0}

\noindent where $M$ refers to the motor. This strongly resembles the
Michaelis--Menten mechanism. The only difference, in this case, is that 
transition rates depend on the coordinate of the protein center of mass. 
We emphasize that all of the following discussion does not involve any 
hypothesis regarding enzymatic catalysis.

The state $M\cdot ADP \cdot P $ corresponds to one in which the motor head 
is detached from the fiber after binding and dissociating ATP. In this state
the motor may move more or less freely upon the filament. This state will be 
called the ``free'' state,  or state $2$. All the other terms in the equations 
refer to a state in which the motor is attached to the filament and therefore 
its motion is strongly dependent on the motor--filament interaction. This is 
the so called ``bound'' state or state $1$. In this case we are dealing 
essentially with a single--headed motor, although the mechanism for a two--head  
motor protein is similar. We assume detailed balance to hold for each chemical 
reaction:

\begin{equation}
\frac{\alpha_1(x)}{\alpha_2(x)} = \exp \left( \frac{W_1(x)-W_2(x)+\Delta
\mu}{k T} \right)
\label{eq:alpha} 
\end{equation}

\begin{equation}
\frac{\beta_1(x)}{\beta_2(x)} = \exp \left(\frac{W_1(x)-W_2(x)}{k T}\right)
\label{eq:beta}
\end{equation}

\noindent where $\alpha_1, \alpha_2\ (\beta_1,\beta_2)$ are the transition rates 
for the first (second) chemical
reaction and $\Delta \mu = \mu_{ATP} - \mu_{ADP} - \mu_P $ is the difference 
in chemical potential or the  chemical driving force.

The FP equation describing the process may now be written as:

\begin{equation}
\partial_t P_i + \partial_x J_i = \sum_{j \neq i} \left( \omega_{ji}
P_j-\omega_{ij} P_i \right)
\end{equation}

\noindent where $i=1,2$ for the two-state model and $\omega_{ij}(x) = 
\alpha_i(x) + \beta_i(x)$.

The main advantage of this approach is the
straightforward manner in which external forces, if any, may be directly inserted in 
eq.~(\ref{eq:fp}) without any need for further assumptions (for a discussion of
this point see~\cite{fish99}). 
If the external force is independent of time, it is possible to look for a 
stationary solution. 
The velocity of the motor, $v$, defined as:

\begin{equation}
v = \int_0^p \left( J_1(x)+J_2(x) \right) dx,
\label{eq:vel}
\end{equation}

\noindent and the rate of ATP consumption:
\begin{equation}
r = \int_0^p \left(\alpha_1(x)P_1(x)-\alpha_2(x)P_2(x)\right) dx ,
\label{eq:rate}
\end{equation}

\noindent where $p$ is the periodicity of  the  filament, are identically zero 
when both $\Delta \mu $ and $F_{ext}$ are zero~\cite{jap97}. A necessary condition for
motion to occur~\cite{jap97} is that $\Delta \mu \neq 0$, i.e. detailed balance
is violated, and that the potential is asymmetric. This implies that the 
chemical reaction of ATP hydrolysis is able to break detailed balance for the 
total transition rates.

Using eq.~(\ref{eq:alpha}) and~(\ref{eq:beta}) only two of the four functions 
$\alpha_1(x), \alpha_2(x), \beta_1(x),$ and $\beta_2(x)$, can be chosen 
arbitrarily once $W_1$ and $W_2$ are fixed. Since the release of products is 
just a thermal process and does not involve chemical reactions (see
eq.~(\ref{eq:adp})), we assume it to be position independent as in~\cite{par99}, 
so that $\beta_2(x)= \omega = const$. 

Transitions due to chemical reactions are usually chosen to be localized, 
i.e. they may take place only in correspondence to a certain motor position 
along the filament period and therefore they are not distributed over the whole period. 
This corresponds to the ``active site'' concept in biology and 
in~\cite{jap97} the former hypothesis is shown to agree with experimental data. 
To take this effect into account we also define:

\begin{equation}
\alpha_2(x) = 
\left\{
\begin{array}{ll}
\omega & \mbox{for } p-\delta < x < p, \\
0 & \mbox{otherwise.}
\end{array}
\right.
\end{equation}

\noindent with $\delta /p = 0.05$.
We use two kinds of models, differing only in the choice of
the potential in state $2$~\cite{jap97,par99}. $W_1$ is the standard ratchet potential shown in
fig.~\ref{fig:model}. In model $(a)$, we suppose state 
$2$ to be strictly diffusive so that the potential $W_2(x)$ is flat 
(Fig.\ref{fig:model}). This corresponds to 
the picture of state $2$ as a totally free state. In model (b), we suppose the 
filament to affect the protein 
movement, so that the potential in state $2$ is essentially the same as
$W_1(x)$, except for a uniform offset and a 5 times smaller amplitude of 
variation. We expect the motion in model (b) to be more constrained than in 
model (a). 

\begin{figure}
\begin{center}
\epsfig{figure=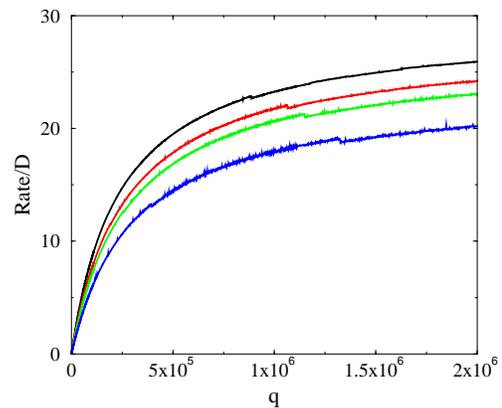,width=5.5cm,angle=270}
\end{center}
\caption{Results for the rate of ATP consumption for model $(a)$ with 
$\omega/D = 50$ and various forces 
(top to bottom curves: $F_{ext}=0.0,-1.0, -3.0, -5.0$ in units of $kT/p$.}
\label{fig:rate}
\end{figure}

We analyzed the two models $(a)$ and $(b)$ seeking the non--equilibrium 
stationary  solutions $P_i^s(x)$ to the FP equations when applying 
different mechanical and chemical driving forces.
This has been accomplished by numerically integrating the stationary FP
eq.~(\ref{eq:fp}) starting from the equilibrium Boltzmann solutions and changing
smoothly and alternatively the parameters $\Delta \mu$ and $F_{ext}$.
For each stationary solution we calculated the velocity and the rate of ATP 
consumption. Fig.~\ref{fig:rate}
shows the results for the rate of ATP consumption.

These results are plotted in terms of the ratio:

\begin{equation}
q=\frac{[ATP]}{[ADP][P]} = \exp \left( \frac{\Delta \mu}{k T} \right)
\label{eq:q}
\end{equation}

\noindent from mass--action law, using concentrations normalized with respect to
their equilibrium value.
Usually the Michaelis law is written in terms of ATP concentration, while in 
our results we derive it 
in terms of the parameter $q$ defined above. We observe that experiments are usually 
performed in conditions of high ATP concentrations, so that the ratio $q$ can be safely 
thought to be constant during the time taken by the experiment. Since the 
only way of varying $q$ is by adding ATP molecules to the solution, $q$ or 
$[ATP]$ may be interchanged in the Michaelis law.
The rate of ATP consumption can be fitted by a Michaelis law in the form:

\begin{equation}
r = \frac{q \cdot r_{MAX}}{K_M + q}.
\label{eq:mich}
\end{equation}

The value $r_{MAX}$ is weakly dependent on the applied force. 
 Our calculations show that the Michaelis--Menten law is followed with a rather impressive
precision for $q\gtrsim 10^3$, where $1/r$ is linear in $1/q$ with regression
coefficients practically equal to $1$.
The fact that the Michaelis law is 
so strictly followed also when different forces are applied is a good 
indication that the simplified two--state model is still in accordance with 
well established results for enzymatic catalysis.

\begin{figure}
\begin{center}
\epsfig{figure=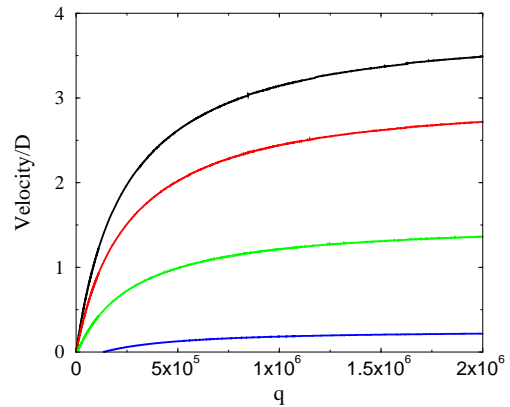,width=5.5cm,angle=270}
\end{center}
\caption[Figure~\ref{fig:vel}]{Results for the sliding velocity for model $(a)$
in units of $p$ with the same values of the forces as in fig.\ref{fig:rate}.}
\label{fig:vel}
\end{figure}

Figure~\ref{fig:vel} shows the velocity as a function of $q$; at first sight
these curves seem Michaelis-like, so that the correct equation for
velocity should be eq.~(\ref{eq:mich}) with $r_{MAX}$ replaced by $v_{MAX}$. 
In this 
case $v_{MAX}$ strongly depends on the applied force, and it decreases when 
the applied force becomes more and more negative. 
Our results show that the velocity
is linear with $r$, so that $v=\alpha r - \beta $ with a regression coefficient 
always $\gtrsim 0.9996$
and thus the
relation for velocity, using eq.~(\ref{eq:mich}), reads:
 
\begin{equation}
v= \frac{\alpha r_{MAX} \cdot q}{K_M+q}-\beta.
\label{eq:michvel}
\end{equation}

A pleasing feature is that, using this equation, there is only one Michaelis 
constant for both $v$ and $r$. Another very interesting point is that the
obtained positive values of $\beta $ indicate that ATP consumption is even
possible under stall conditions ($v=0$) \cite{ref}. 
This ``idling'' rate of ATP consumption should be tested by motility assays.

The Michaelis law is commonly thought to hold for very high ATP concentrations. 
The experimental results~\cite{schn99} were obtained in this regime and fitted
using eq.~(\ref{eq:michvel}) forcing $\beta = 0$. Table~\ref{t:res} shows the
results obtained by fitting our numerical data using eq.~(\ref{eq:michvel}) 
with $\beta $ as a free parameter; these results are compared with the ones
obtained in a fit with a standard Michaelis law, i.e. forcing $\beta = 0$.
In the former case the fit is extremely good; indeed the sum
of square residuals can be hundreds times less than in the latter case.
While the maximum attained velocity is essentially independent of the fitting,
the results for the Michaelis constant are quite different. 
The range of variability of $K_M$ with $F_{ext}$ is much wider using a
standard Michaelis law; this is also confirmed
by simple analytical arguments. Furthermore, in the case of model $(b)$ for high
loads  it is even impossible to fit the data without introducing $\beta $ 
in eq.~(\ref{eq:michvel}).
Still, eq.~(\ref{eq:exp}) does not allow for any inversion in the sign of 
velocity, nor for a stall condition, at variance with eq.~(\ref{eq:michvel}).

In conclusion, our calculations on the two-state ratchet model 
clearly show that the rate of ATP consumption strictly follows a Michaelis 
law in the form~(\ref{eq:mich}); that the correct law for velocity is given by
eq.~(\ref{eq:michvel}), instead of eq.~(\ref{eq:exp}); that the observed fourfold
increase of the Michaelis constant with applied load may be due to the use of
eq.~(\ref{eq:exp}); that the model still accounts for an increase of the 
Michaelis constant smaller than $20\%$ when forces vary by a factor $5$ 
(in the range of pN), to be compared with the corresponding variation of about
$250\%$ if a standard Michaelis  law, eq.~(\ref{eq:exp}), is used.

We thank F.~Cecconi and P.~V\'an for useful discussions. GL also acknowledges 
INFM funding.


\begin{table}
\begin{center}
\begin{tabular}{|r|r|r|r|r|} \hline 
 $F_{ext}$ &  $K_M\ \ $  & $K'_M$ & $ v_{MAX}$ & $v'_{MAX}$\\ \hline
 $  0.0  $ & $238030$ & $243710$ & $3.9053$ & $3.9067$ \\ \hline
 $ -1.0  $ & $249970$ & $254680$ & $3.079$  & $3.0573$ \\ \hline
 $ -2.0  $ & $263440$ & $268840$ & $2.316$  & $2.2726$ \\ \hline
 $ -3.0  $ & $272320$ & $292800$ & $1.584$  & $1.5541$ \\ \hline
 $ -4.0  $ & $282930$ & $360000$ & $0.9453$ & $0.899$ \\ \hline
 $ -5.0  $ & $296040$ & $\sim 900000$ & $0.3886$ & $0.295$ \\ \hline
\end{tabular}
\end{center}
\caption{ \label{t:res} Michaelis constants, $K_M$, and maximum velocities,
$v_{max}=\alpha \ r_{max}$ calculated
from model $(a)$ using eq.~(\ref{eq:michvel}) (second and fourth columns) with
$\beta $ as a free parameter and the same equation with $\beta = 0 $ (third and
fifth columns) corresponding to the standard Michaelis law. In the latter case,
the constants $K_M$ and $v_{max}$ are indicated as $K'_M$ and $v'_{max}$
respectively. Forces (first column) are measured in units of $kT/p$.} 
\end{table}

\end{document}